# Ferroelectricity in free standing perovskite-based nanodots: A density functional theory study

Karthik Guda Vishnu,[1] Samuel Temple Reeve[2], and Alejandro Strachan[1*]

[1]School of Materials Engineering and Birck Nanotechnology Center, Purdue University, West Lafayette, Indiana 47906

[2]Materials Science Division, Lawrence Livermore National Laboratory, Livermore, California, USA 94550

## Abstract:

We use density functional theory to investigate the possibility of polar and multiferroic states in free-standing, perovskite-based nanodots at their limit of miniaturization: single unit cell with termination to allow centrosymmetry. We consider both A-O and B-$O_2$ terminations for three families of nanodots: i) A=Ba with B=Ti, Zr, and Hf; ii) A=Ca and Sr with B=Ti; and iii) A = Na and K with B=Nb. We find all A-O terminated dots to be non-polar and to exhibit cubic symmetry, regardless of the presence of ferroelectricity in the bulk. These dots all carry a net magnetic moment, except $Ca_8TiO_6$. On the other hand, all B-$O_2$ terminated nanodots considered in this study relax to a non-cubic ground state. Rather surprisingly, a subset of these structures ($BaTi_8O_{12}$, $BaHf_8O_{12}$, $BaZr_8O_{12}$, $SrTi_8O_{12}$, $CaTi_8O_{12}$ and $KNb_8O_{12}$) exhibit polar ground states. We propose a new structural parameter, the cluster tolerance factor (CTF), to determine whether a particular combination of A and B will yield a polar ground state in the nanodot geometry, analogous to the Goldschmidt factor in bulk ferroelectrics. We also find that in all polar B-$O_2$ terminated nanodots are also magnetic. The multiferroic state originates from separation between spin density in peripheral B atoms and polarity primarily caused by the off-center central A atom. Our findings stress that surface termination plays a crucial role in determining whether ferroelectricity is completely suppressed in perovskite-based materials at their limit of miniaturization.

# Introduction:

Ferroelectricity in bulk materials is dominated by the competition between long-range electrostatics and short-range repulsive forces (Pauli exclusion).[1] The former is key in the stabilization of the polar ferroelectric phase while the latter tends to stabilize the centrosymmetric, non-polar cubic phase. As such, it is not surprising that complex size effects arise when bulk ferroelectrics are miniaturized, as long-range interactions are disproportionally affected. Numerous factors affect ferroelectricity in nanoscale materials: the geometry of the system, the material chemistry, the nature of the boundary conditions, and the nature of surface termination/passivation.[2,3,4,5,6,7,8] Understanding these effects and ferroelectricity at the limit of miniaturization has gathered significant interest in recent years, as nanoscale ferroelectrics[9] are of current interest in non-volatile memory devices,[10,11,12,13,14] nanoscale actuators, and capacitive transducers.[15,16] Furthermore, nanoscale systems have an inherent non-stoichiometry which can result in uncompensated spin carriers and magnetism. The coexistence of ferroelectricity and magnetism is rare and prior to the advent of multiferroics, these properties were thought to be mutually exclusive:[17,18,19,20] magnetism requires unpaired electrons in the d-shell, while ferroelectricity is promoted by the d-shell hybridization. However, in materials like $BiFeO_3$ and $BiMnO_3$, the most popular multiferroics, simultaneous ferromagnetic ordering and ferroelectricity is achieved by isolating the ferroelectricity to the Bi atom (polarization is caused by the off-center displacements of Bi atom) and ferromagnetism to the B cation (Fe, Mn).[17,18] A recent density functional theory (DFT) study on $PbTiO_3$ nanodots[7] revealed a complex interplay between ferroelectricity and ferromagnetism as a function of their size and the nature of their surface termination leading to multiferroic properties. The ability to stabilize such multiferroic phases in perovskite-based nanodots is of great interest and an understanding of the role of different chemistries and surface terminations at their limit of miniaturization could lead to potential novel device design concepts.[21,22]

In this paper, we use DFT using accurate hybrid functionals to characterize the multiferroic properties of a wide range of free-standing, perovskite based ($ABO_3$), nanodots at their limit of miniaturization: a single unit cell with the two possible terminations that support cubic symmetry, shown in Figure 1. We find that the ground states for A-O terminated dots and chemical formula $A_8BO_6$, exhibit cubic symmetry. However, for dots with B-$O_2$ termination and chemical formula $AB_8O_{12}$, non-cubic ground states are the rule and not the exception. More importantly, all B-$O_2$ terminated nanodots except $NaNb_8O_{12}$, are not only ferroelectric but also exhibit, magnetism.

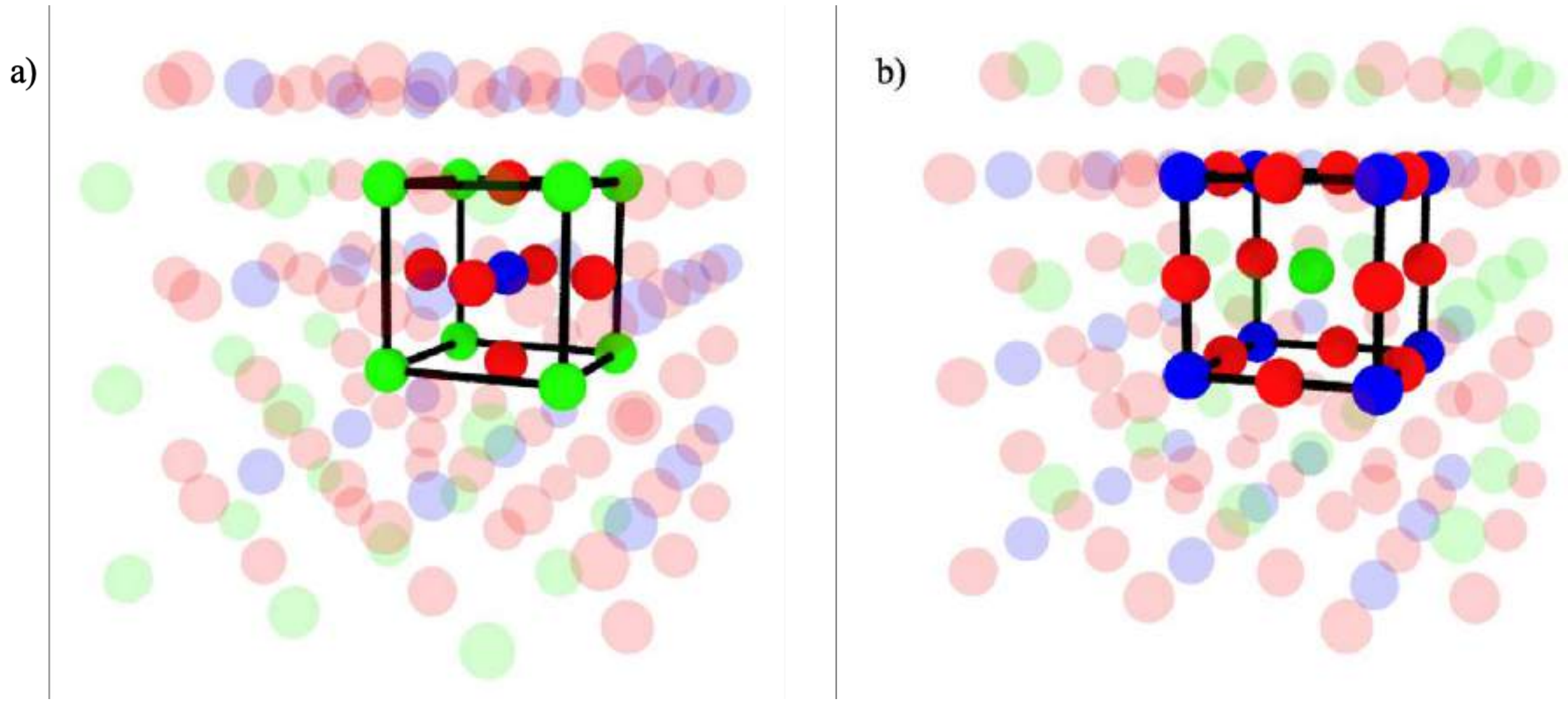

*Figure 1: Illustration of (a) A-O terminated nanodot with chemical formula $A_8BO_6$ and (b) B-$O_2$ terminated nanodot with chemical formula $AB_8O_{12}$. A atoms are in green, B atoms in blue and O atoms in red. Black lines connecting the atoms in the nanodots highlight a single unit cell.*

The remainder of this paper is organized as follows. In Section 2 we provide the simulation details. Section 3 shows results on polarization and discussion of the nanodot ground state symmetries, followed by a description of the observed multiferroic nature of the various polar nanodots. The proposed cluster tolerance factor (CTF) is explained and bond length analysis is presented in detail in Section 4. Finally, in Section 5 we present our summary and conclusions.

## 2. Computational details

We simulate both A-O and B-$O_2$ terminated nanodots for a wide range of A and B combinations: i) A = Ba with B = Ti, Hf, and Zr; ii) A = Ca and Sr with B = Ti; and iii) A = K and Nb with B = Nb. As shown in Figure 2(a) the A-O terminated nanodots contain 15 atoms with the B atom at the center of the structure enclosed by a shell of A and O atoms, with chemical formula $A_6BO_8$. Similarly, Figure 2(b) shows the 21-atom B-$O_2$ terminated nanodots, with the A atom at the center enclosed by a B shell and an O shell and with chemical formula $AB_8O_{12}$. All visualization was performed using VESTA.[23]

## 2.1 DFT codes and details

In order to obtain accurate ground state structures and their associated properties, we follow a two-step process which includes an initial relaxation using SeqQuest,[24,25,26] a DFT[27] code developed at Sandia National Laboratories, within the generalized gradient approximation (GGA) of Perdew, Burke, and Ernzerhof (PBE).[28] This is step is referred to as the ***PBE-SeqQuest*** method throughout the manuscript. The relaxed structures obtained from this first step are re-relaxed using the ORCA code[29,30] and the B3LYP (Becke, 3-parameter Lee-Yang-Parr) hybrid functional.[31,32] This step is referred to as the ***B3LYP-ORCA*** method throughout the manuscript. Although DFT simulations with PBE have previously been used to accurately describe the bulk lattice parameters, elastic constants, surface energies, and surface atomic relaxations in various perovskite oxides[33], this two-step approach is important as semi-local DFT methods tend to underestimate bandgaps and magnetic moments in systems containing transition metal atoms with partially filled d and f shells.[34,35] Further simulation details of both the steps are provided below.

***PBE-SeqQuest*** SeqQuest uses contracted Gaussian functions as a basis set; our calculations were performed using double-zeta polarized basis sets. Norm conserving pseudopotentials of the Hamann type,[36] parametrized for the PBE functional, were used to describe core electrons. All calculations were spin polarized and carried out at 0.04 eV electronic temperature. SeqQuest uses the maximum change in any Hamiltonian matrix element as its self-consistent field (SCF) convergence criterion; this was set to $6.8 \times 10^{-2}$ eV for all calculations. SeqQuest uses local moment counter charge (LMCC) method[37] to exactly eliminate any spurious dipole interactions. Real space grid spacing was set at 0.13 Å in all simulations. We used only gamma point to sample the reciprocal space as all our systems are non-periodic in all directions. Different spin states were investigated for each configuration and no anti-ferromagnetic configurations were considered in this study. Supercells containing the nanodots also included at least 10 Å of vacuum along x, y and z directions. Each nanodot structure was fully relaxed with respect to atomic positions starting from various initial structures using the accelerated steepest descent (ASD) method.[24] Convergence was assumed when the absolute force on every atom was less than or equal to $25.7 \times 10^{-3}$ eV/Å.

***B3LYP-ORCA*** B3LYP hybrid functional as implemented in ORCA was used to re-relax the structures obtained using SeqQuest. A standard B3LYP functional with 20% Hartree-Fock (HF) exchange was used. ORCA uses effective core potentials (ECPs)[38,39] for all atoms heavier than Kr. We used Ahlrich's valence triple-zeta polarized basis set called def2-TZVP[40] in all simulations. The charge on each structure was set to zero, while the ground state spin state was chosen by running the geometric optimization at various spin multiplicities and choosing the lowest energy structure.

## 2.2 Structure relaxation, symmetry, and analysis

***PBE-SeqQuest*** We first simulated the high-symmetry, paraelectric structure for all cases by initializing the atomic positions corresponding to the cubic structure and restricting the symmetry of the nanodot during the relaxation to remain in the $O_h$ (cubic) point group. In order to explore the possibility of non-cubic and polar ground state structures, we also initialized the relaxations from dots with symmetry broken along $[001]_{cub}$ (tetragonal), $[110]_{cub}$ (orthorhombic), and $[111]_{cub}$ (rhombohedral). We did not impose any symmetry restriction during the atomic relaxation in the above three configurations. We evaluated the formation energies ($E_{form}$) of these nanodots based on the ground state crystal structures of their constituent elements:

$$E_{form} = \frac{E_{tot} - (n_A\mu_A + n_B\mu_B + n_O\mu_O)}{n_A + n_B + n_O}$$

where $E_{tot}$ is the total energy of the nanodot; $\mu_i$ is the chemical potential of element i in its standard form, the ground state crystal structure for the A and B atoms and the $O_2$ molecule in the triplet state; and $n_A$, n2, and n3 are the number of A, B, and O atoms, respectively. The total dipole moment for all relaxed structures was obtained from the relaxed electronic density in the DFT calculation:

$$\vec{P} = -e\int \vec{r}\rho(\vec{r})d^3\vec{r}$$

We use dipole moments obtained from electronic densities throughout for the PBE-SeqQuest simulations. A comparison between the dipole moments obtained from the electronic densities and partial atomic charges can be found in the supplementary material, Section S1, for completeness.

***B3LYP-ORCA*** The ground state structures obtained using PBE-SeqQuest method were further relaxed using the B3LYP hybrid functional available in ORCA. No symmetry constraints were imposed while determining the ground state structures. For the cubic non-polar nanodots, in order to fix the symmetry to $O_h$ point group, we adopt a constrained optimization procedure where we fix all the angles and dihedrals starting from relaxed cubic non-polar ($O_h$) structures obtained using PBE-SeqQuest. Due to the non-periodic nature of the ORCA DFT code, $E_{form}$ defined in equation (1) cannot be evaluated and hence only the relative energy difference with respect to cubic non-polar dots is reported. ORCA internally calculates the dipole moment as the negative derivative of the energy with respect to an external electric field, determined as the expectation value of the variational wave function. Total dipole moments obtained as a sum of electronic and nuclear contributions are used throughout for the B3LYP-ORCA simulations.

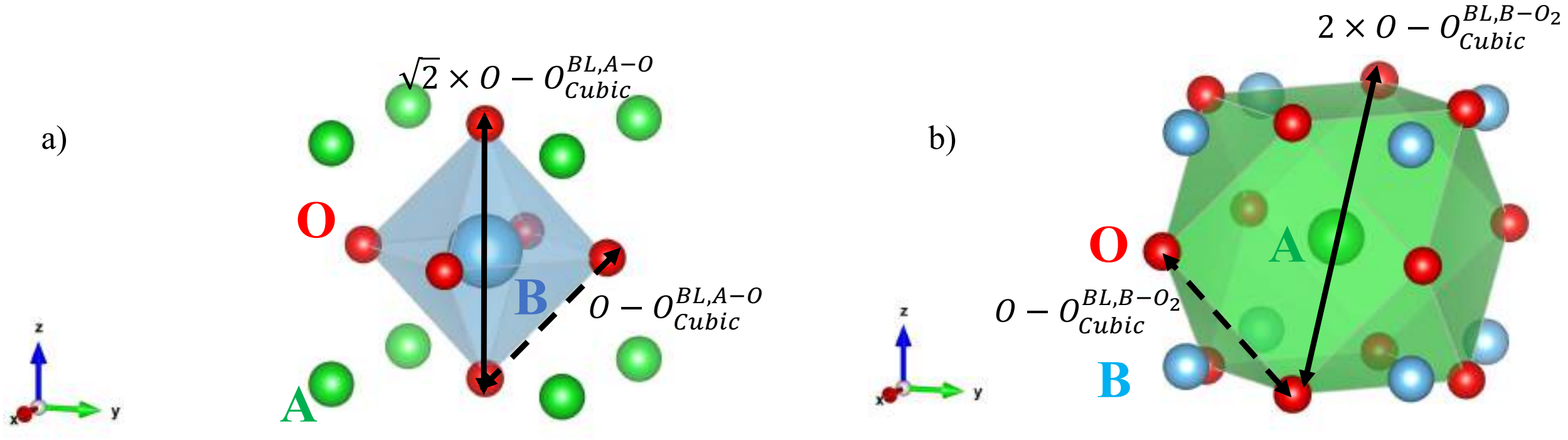

*Figure 2: Cubic structure, with a polyhedron surrounding the central atom of a) an A-O terminated nanodot and b) a B-$O_2$ terminated nanodot. For a description about solid and dashed lines, see section 4.*

# 3. Ground state structures and multiferroic properties

## 3.1 Energetics, symmetry, and electrical polarization

Tables 1 and 2 summarizes energetics, key structural properties, and dipole moments of the ground state configuration for each dot studied obtained using both the PBE-SeqQuest and B3LYP-ORCA methods. As explained in Section 2, due to the inherent advantages of B3LYP-ORCA method, we focus on the results obtained using the B3LYP-ORCA method throughout the rest of the manuscript and we use results from PBE-SeqQuest method for comparison only where necessary.

For all A-O terminated nanodots, the ground state is a cubic structure with $O_h$ point group, except for $K_8NbO_6$. The $K_8NbO_6$ nanodot relaxes to a non-polar non-cubic structure with $D_{2h}$ symmetry, whose energy is lower than its cubic $O_h$ counterpart by 2 meV/atom. Furthermore, all A-O dots exhibit large magnetic moments except for $Na_8NbO_6$ and $K_8NbO_6$, as well as $Ca_8TiO_6$ which relaxes to a structure with zero net magnetic moment. Importantly, we find that ferroelectricity is completely suppressed at the limit of miniaturization in the A-O terminated nanodots for all chemistries considered in this study. Our results are consistent with the latest predictions on 1 unit-cell thick 2-dimensional $BaTiO_3$ slabs made using effective Hamiltonian based finite temperature simulations.[41]

Quite remarkably, all B-$O_2$ terminated nanodots, relax to non-cubic ground state structures and all of these structures exhibit electric dipole moments, except for $NaNb_8O_{12}$. Symmetry analysis shows both $NaNb_8O_{12}$ nanodot relaxes to a non-cubic, centrosymmetric (i.e. non-polar) structure with the point group $C_{2h}$, whose lowest non-vanishing multipole moment is a quadrapole moment. The $BaTi_8O_{12}$, $BaZr_8O_{12}$, and $KNb_8O_{12}$ nanodots relax to $C_1$ configuration with no symmetry and with dipole magnitudes of 1.29 D, 4.92 D and 5.53 D respectively. While $BaTi_8O_{12}$ nanodot has no preferred axial dipole orientation, both $BaZr_8O_{12}$ and $KNb_8O_{12}$ nanodots exhibit preferential dipole moment along their z-direction ($[001]_{cubic}$). Ground state structures of $SrTi_8O_{12}$ and $BaHf_8O_{12}$ nanodots are polar with point group $C_{2v}$, within the orthorhombic class; consistent with this symmetry we find that both nanodots exhibit dipole moments only in the $(011)_{cubic}$ planes. On the other hand, $CaTi_8O_{12}$ nanodot relaxes to a non-cubic polar ground state structure belonging to the point group $C_s$ with significant dipole moments along all three principal axes ($<100>_{cubic}$). Both $SrTi_8O_{12}$ and $KNb_8O_{12}$ carry the highest net dipole moments (7.77 D and 5.52 D, respectively) among all systems investigated. All B-$O_2$ terminated nanodots also carry a net spin, making them multiferroic, except for $NaNb_8O_{12}$ whose dipole moment is zero. Snapshots of all the non-cubic ground state nanodot structures obtained from B3LYP-ORCA are shown in Figure 3. It is clear from Figure 3 that in all polar nanodots off-center displacement of the central A atom is the majority contributor to the overall polarity of the nanodot.

It is clear from our results that the existence of a polar ground state in the bulk does not guarantee a polar ground state in the nanodot geometry considered in this study, as evidenced by $NaNbO_3$. More interestingly, the absence a polar bulk phase does not preclude polarization in the nanodots as shown in the case of $SrTiO_3$, $BaZrO_3$, $CaTiO_3$ and $BaHfO_3$. Finally, the case of $KNbO_3$ highlights the case where polarization stabilized in both the nanodot geometry and the bulk.

From our PBE-SeqQuest calculations, we find that the ground state B-$O_2$ terminated nanodots have significantly lower formation energies compared to their A-O terminated counterparts for all cases, indicating greater stability. PBE-SeqQuest similarly predicts that all A-O terminated nanodots relax into non-polar cubic structures and also matches the B3LYP-ORCA for B-$O_2$ systems, with the exception of $BaTi_8O_{12}$. More detailed discussion regarding the energetics, symmetry, and electrical polarization of the PBE-SeqQuest nanodots can be found in supplementary material, Sections S2-3.

## 3.2 Magnetism and presence of multiferroic states

We now focus on the origin of multiferroic states in the nanodots. Figure 4 (a-f) shows the spin density distribution for various polar nanodots together with those of their respective cubic structures. It is clear from Figure 4, that in all polar nanodots the maximum spin density carrier are the peripheral B atoms while the polarity is caused primarily by the off-center displacements of the central A atom. As mentioned before, this phenomenon is similar to the isolation mechanism observed in bulk multiferroics, such as $BiFeO_3$ and $BiMnO_3$. Further, in all polar nanodots, the overall magnetic moment decreases from the cubic, non-polar structure to the relaxed, polar structure. Two exceptions are $BaZr_8O_{12}$ (see Figure 4(b)), where the magnetic moment remains constant, and $BaHf_8O_{12}$ (see Figure 4(e)), where the magnetic moment increases. Our PBE-SeqQuest calculations predict that only $SrTi_8O_{12}$ and $KNb_8O_{12}$ nanodots assume multiferroic order, as semi-local DFT functionals tend to underestimate magnetism. The mechanism for the stabilization of multiferroicity remains similar to that observed in bulk multiferroics. Full spin density results from the PBE-SeqQuest calculations are shown in the supplementary material, Section S3.

*Table 1: Summary of energetics, dipole moments and spin of all the A-O terminated ground state structures of all nanodots studied. A-Shell characteristic length (CL), O-Shell characteristic length (CL) and CTF are calculated from corresponding cubic non-polar structures. Nanodots with zero dipole moment are classified as nonpolar (NP) and those with non-zero dipole moment are classified as polar (P).*

| Nanodot | Method | $E_{form}$ (eV/ atom) | $\Delta E_{cub-sym}$ (eV/ atom) | Point group | Dipole moment (Debye) | | | | Majority Spin | A Shell CL (Å) | O Shell CL (Å) | CTF | Nanodot polarity (P/NP) |
|---|---|---|---|---|---|---|---|---|---|---|---|---|---|
| | | | | | X | Y | Z | Total | | | | | |
| $Ba_8TiO_6$ | PBE-SeqQuest | -1.351 | - | $O_h$ | 0.0 | 0.0 | 0.0 | 0.0 | 6.0 | 3.828 | 4.044 | 1.008 | NP |
| $Ba_8TiO_6$ | B3LYP-ORCA | - | - | $O_h$ | 0.0 | 0.0 | 0.0 | 0.0 | 8.0 | 3.802 | 4.017 | 1.002 | NP |
| $Ba_8ZrO_6$ | PBE-SeqQuest | -1.320 | - | $O_h$ | 0.0 | 0.0 | 0.0 | 0.0 | 6.0 | 3.91 | 4.244 | 1.001 | NP |
| $Ba_8ZrO_6$ | B3LYP-ORCA | - | - | $O_h$ | 0.0 | 0.0 | 0.0 | 0.0 | 8.0 | 3.901 | 4.201 | 0.991 | NP |
| $Ba_8HfO_6$ | PBE-SeqQuest | -1.373 | - | $O_h$ | 0.0 | 0.0 | 0.0 | 0.0 | 6.0 | 3.905 | 4.182 | 0.991 | NP |
| $Ba_8HfO_6$ | B3LYP-ORCA | - | - | $O_h$ | 0.0 | 0.0 | 0.0 | 0.0 | 8.0 | 3.908 | 4.186 | 0.992 | NP |
| $Ca_8TiO_6$ | PBE-SeqQuest | -1.364 | - | $O_h$ | 0.0 | 0.0 | 0.0 | 0.0 | 6.0 | 3.39 | 4.033 | 1.006 | NP |
| $Ca_8TiO_6$ | B3LYP-ORCA | - | - | $O_h$ | 0.0 | 0.0 | 0.0 | 0.0 | 0.0 | 3.368 | 4.006 | 0.999 | NP |
| $Sr_8TiO_6$ | PBE-SeqQuest | -1.418 | - | $O_h$ | 0.0 | 0.0 | 0.0 | 0.0 | 6.0 | 3.579 | 4.032 | 1.005 | NP |
| $Sr_8TiO_6$ | B3LYP-ORCA | - | - | $O_h$ | 0.0 | 0.0 | 0.0 | 0.0 | 8.0 | 3.618 | 3.972 | 0.991 | NP |
| $Na_8NbO_6$ | PBE-SeqQuest | -1.242 | - | $O_h$ | 0.0 | 0.0 | 0.0 | 0.0 | 0.87 | 3.28 | 4.108 | 1.007 | NP |
| $Na_8NbO_6$ | B3LYP-ORCA | ---- | --- | $O_h$ | 0.0 | 0.0 | 0.0 | 0.0 | 1.0 | 3.239 | 4.058 | 0.995 | NP |
| $K_8NbO_6$ | PBE-SeqQuest | -1.202 | - | $O_h$ | 0.0 | 0.0 | 0.0 | 0.0 | 0.34 | 3.761 | 4.133 | 1.013 | NP |
| $K_8NbO_6$ | B3LYP-ORCA | ---- | -0.002 | $D_{2h}$ | 0.0 | 0.0 | 0.0 | 0.0 | 1.0 | 3.704 | 4.07 | 0.998 | NP |

*Table 2: Summary of energetics, dipole moments and spin of all the B-$O_2$ terminated ground state structures of all nanodots studied. B-Shell characteristic length (CL), O-Shell characteristic length (CL) and CTF are calculated from corresponding cubic non-polar structures. Nanodots with zero dipole moment are classified as nonpolar (NP) and those with non-zero dipole moment are classified as polar (P).*

| Nanodot | Method | $E_{form}$ (eV/ atom) | $\Delta E_{cub\text{-}sym}$ (eV/ atom) | Point group | Dipole moment (Debye) | | | | ($N_\alpha$-$N_\beta$) | B Shell CL (Å) | O Shell CL (Å) | CTF | Nanodot polarity (P/NP) |
|---|---|---|---|---|---|---|---|---|---|---|---|---|---|
| | | | | | X | Y | Z | Total | | | | | |
| $BaTi_8O_{12}$ | PBE-SeqQuest | -2.082 | - | $O_h$ | 0.0 | 0.0 | 0.0 | 0.0 | 10.0 | 3.661 | 4.242 | 0.997 | NP |
| $BaTi_8O_{12}$ | B3LYP-ORCA | | -0.145 | $C_1$ | -0.48 | 0.96 | 0.73 | 1.29 | 8.0 | 3.637 | 4.214 | 0.99 | P |
| $BaZr_8O_{12}$ | PBE-SeqQuest | -2.033 | -0.099 | $C_{4v}$ | 0.01 | -0.02 | -4.97 | 4.97 | 0.0 | 3.627 | 4.86 | 1.142 | P |
| $BaZr_8O_{12}$ | B3LYP-ORCA | | -0.124 | $C_1$ | 0.59 | -0.21 | 4.88 | 4.92 | 4.0 | 3.618 | 4.848 | 1.139 | P |
| $BaHf_8O_{12}$ | PBE-SeqQuest | -2.072 | -0.047 | $C_{4v}$ | 0.04 | 0.03 | -5.69 | 5.69 | 0.0 | 3.603 | 4.745 | 1.115 | P |
| $BaHf_8O_{12}$ | B3LYP-ORCA | | -0.161 | $C_{2v}$ | -0.05 | 0.0 | 4.80 | 4.80 | 6.0 | 3.658 | 4.817 | 1.132 | P |
| $CaTi_8O_{12}$ | PBE-SeqQuest | -2.186 | -0.132 | $C_{2h}$ | 0.0 | 0.0 | 0.0 | 0.0 | 2.0 | 3.217 | 4.512 | 1.164 | P |
| $CaTi_8O_{12}$ | B3LYP-ORCA | | -0.148 | $C_s$ | 1.24 | 1.32 | 2.67 | 3.22 | 2.0 | 3.225 | 4.524 | 1.168 | P |
| $SrTi_8O_{12}$ | PBE-SeqQuest | -2.135 | -0.045 | $C_s$ | -0.06 | 4.87 | 4.13 | 6.38 | 4.0 | 3.587 | 4.229 | 1.053 | P |
| $SrTi_8O_{12}$ | B3LYP-ORCA | ---- | -0.05 | $C_{2v}$ | 0.00 | 7.77 | 0.23 | 7.77 | 4.0 | 3.591 | 4.234 | 1.054 | P |
| $NaNb_8O_{12}$ | PBE-SeqQuest | -1.510 | -0.005 | $C_i$ | 0.0 | 0.0 | 0.0 | 0.0 | 0.98 | 3.058 | 4.767 | 1.208 | NP |
| $NaNb_8O_{12}$ | B3LYP-ORCA | | -0.007 | $C_{2h}$ | 0.0 | 0.0 | 0.0 | 0.0 | 1.0 | 3.032 | 4.727 | 1.198 | NP |
| $KNb_8O_{12}$ | PBE-SeqQuest | -1.284 | -0.109 | $C_s$ | 0.44 | 4.79 | -4.85 | 6.83 | 5.03 | 3.206 | 4.814 | 1.120 | P |
| $KNb_8O_{12}$ | B3LYP-ORCA | ------ | -0.187 | $C_1$ | -0.93 | -1.57 | 5.27 | 5.53 | 1.0 | 3.182 | 4.779 | 1.112 | P |

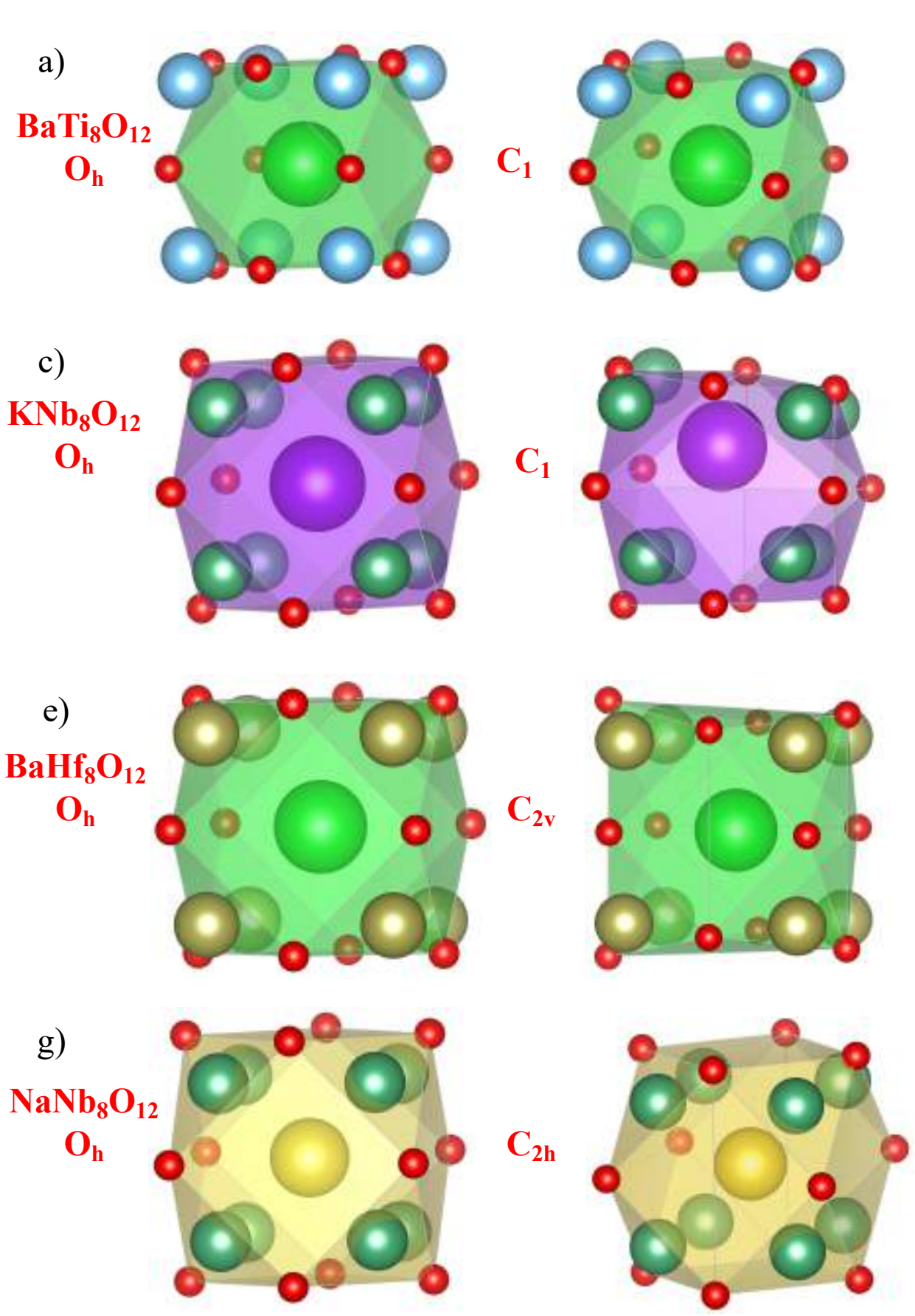

*Figure 3:All $BO_2$ terminated nanodots, with a polyhedron around the central atom, with cubic structures (L) and non-cubic ground state structures (R)*

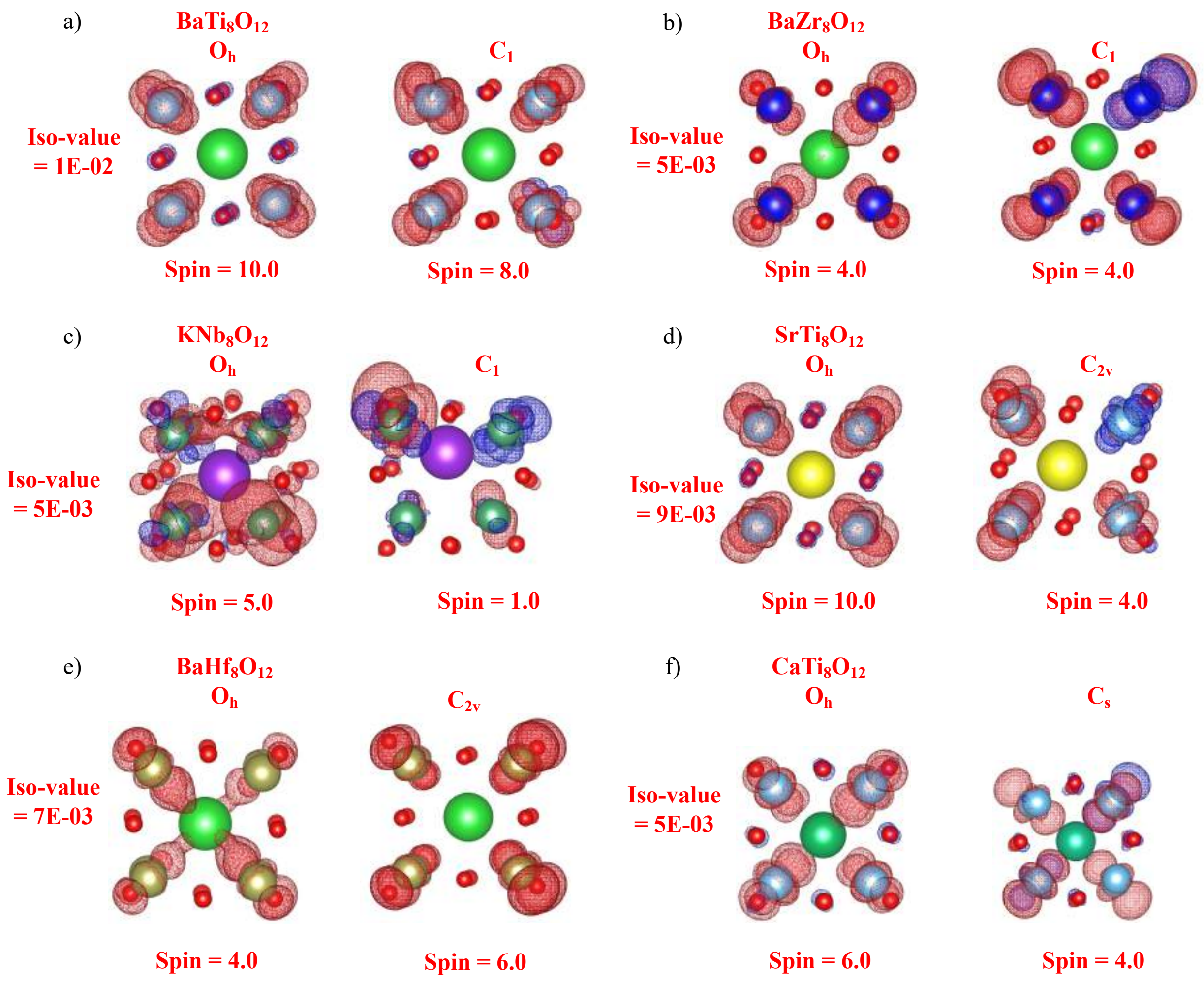

*Figure 4: Spin density ($\rho_{up}$ - $\rho_{dn}$) plots for polar nanodots with their cubic structures (L) and ground state (R). Iso-surface values are noted for each. Red and blue colors indicate positive and negative values, respectively.*

## 4. Cluster tolerance factor

In order to explain the non-trivial nature of the aforementioned results, we propose an empirical criterion for the stability of non-polar, centrosymmetric A-O terminated cubic nanodots and B-$O_2$ terminated cubic nanodots, analogous to Goldschmidt tolerance factor (GTF)[42] used for bulk perovskites to determine the stability of the cubic structure. GTF for bulk perovskites is defined as:

$$GTF = \frac{R_A + R_O}{\sqrt{2}(R_B + R_O)}$$

where $R_A$, $R_B$ and $R_O$ are the atomic radii of elements A, B, and O, respectively. Assuming hard spheres model, GTF measures how well the A, B and O atoms pack the cubic perovskite structure. In an ideal undistorted cubic perovskite with lattice parameter a, the following relationships hold:

$$a = \sqrt{2}(R_A + R_O) = 2(R_B + R_O)$$

thus, making the GTF equal to 1. Hence, it was postulated that when the GTF ~ 1, a cubic, non-polar, centrosymmetric structure is preferred as the ground state and when the GTF deviates from unity, non-cubic structures are preferred as ground states. It should be noted that GTF deviation away from unity likely indicates the stabilization of non-cubic structures as ground states but does not guarantee a polar ferroelectric structure as ground state; for example, $CaTiO_3$ has a GTF value of 0.966, but relaxes to a non-polar, centrosymmetric orthorhombic structure. For all systems considered in this study, the GTF correctly predicts the stabilization of non-cubic structures in bulk $BaTiO_3$, $CaTiO_3$, $KNbO_3$, and $NaNbO_3$ and that bulk $BaZrO_3$, $BaHfO_3$, and $SrTiO_3$ remain cubic.

We propose a cluster tolerance factor (CTF) based on the radius of the central atom, oxygen atom, and the Oxygen – Oxygen bond length of the O-shell (see dotted line in figures 2(a) and 2(b)), which is determined from the relaxed cubic, centrosymmetric and non-polar nanodot structures. Assuming that the close-packed direction for the A-O terminated cubic nanodots is along the $[100]_{O\text{-}Shell}$ (see solid line in Fig 2(a)) and for the B-$O_2$ terminated cubic nanodots it is along $[110]_{O\text{-}shell}$ (see solid line in Fig 2(b)), cluster tolerance factor (CTF) is defined as shown in the equations below:

$$CTF^{A-O} = \frac{\sqrt{2} \times O - O_{Cubic}^{BL,A-O}}{2 \times (R_B + R_O)} = \frac{O - O_{Cubic}^{BL,A-O}}{\sqrt{2} \times (R_B + R_O)}$$

$$CTF^{B-O_2} = \frac{2 \times O - O_{Cubic}^{BL,B-O_2}}{2 \times (R_A + R_O)} = \frac{O - O_{Cubic}^{BL,B-O_2}}{(R_A + R_O)}$$

Shannon ionic radii[43] are used in computing CTF. We postulate that when the CTF ~ 1, cubic non-polar centrosymmetric structure is preferred as the ground state whereas when CTF deviates from 1, the ground state will be a non-cubic structure. We can see from Table 1 that for all A-O terminated nanodots (B3LYP-ORCA) the CTF is approximately 1 (within ±0.01); we accordingly predict that a non-polar centrosymmetric structure is preferred, consistent with the findings of all explicit DFT simulations. For B-$O_2$ terminated nanodots (B3LYP-ORCA) the CTFs are significantly different from unity (at least ±0.05), except for $BaTi_8O_{12}$, also consistent with explicit simulation results (see Table 2). CTF predictions from PBE-SeqQuest calculations are consistent with those of B3LYP-ORCA for all cases.

Further, we calculated the A-O and B-O bond lengths in non-polar centrosymmetric A-O terminated and B-$O_2$ terminated nanodots and compared them against the sum of their ionic radii,[43] as shown in Figure 5. We find that in case of B-$O_2$ terminated nanodots, B-O bond lengths are significantly shorter than the sum of their ionic radii while the A-O bond lengths are comparable to their sum of ionic radii. In the case of A-O terminated nanodots, B-O bond lengths are larger or comparable to their sum of ionic radii while A-O bond lengths are shorter when compared to their sum of ionic radii. This indicates that the B-O atoms in the B-$O_2$ terminated nanodots are likely hybridized, whereas A-O atoms are likely hybridized is in A-O terminated nanodots. This shows that B-O hybridization leads to the stabilization of non-cubic structures in all nanodots, similar to the bulk.[1,44] However, we find that A-O hybridization plays no role in stabilizing non-cubic nanodot structures. Bond length analysis of PBE-SeqQuest relaxed structures yield similar results, as shown in the supplementary material, Section S4.

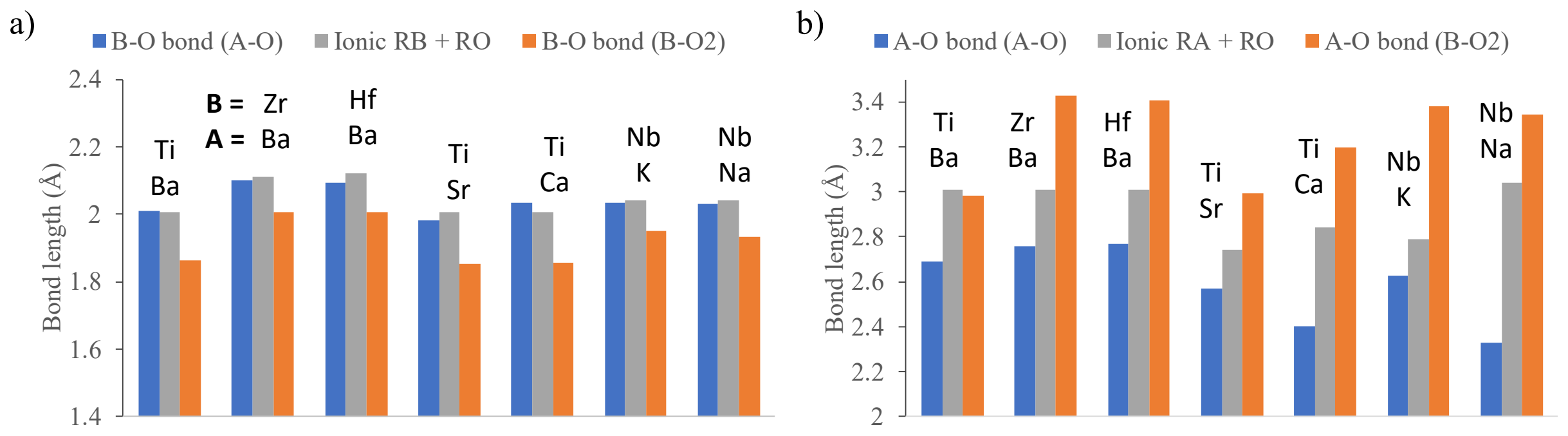

*Figure 5: a) Comparison of B-O bond lengths from both cubic centrosymmetric A-O and B-O2 terminated nanodots against the sum of ionic radii b) Comparison of A-O bond lengths from both cubic centrosymmetric A-O and B-$O_2$ terminated nanodots against their sum of ionic radii*

## 5. Summary and conclusions

In summary, we investigated the existence of polar and multiferroic states at the limit of miniaturization across several families of perovskite-based nanodots in vacuum using explicit hybrid DFT simulations. We find that B-$O_2$ terminated nanodots exhibit lower formation energies than A-O terminated nanodots across all chemistries considered. More importantly, we find that the presence or absence of ferroelectricity in the bulk does not correlate with the properties of the nanodots at the limit of miniaturization.

We find that all A-O terminated nanodots assume a cubic, centrosymmetric, non-polar structure in their ground state (except $K_8NbO_6$), corresponding to the point group $O_h$, while carrying a significant magnetic moment (except for $K_8NbO_6$ and $Na_8NbO_6$), making them neither ferroelectric nor multiferroic, only magnetic. The A-O terminated $Ca_8TiO_6$ nanodot relaxes into a cubic, centrosymmetric, non-polar and also non-magnetic structure making it neither ferroelectric nor magnetic. We find that nanodots in their B-$O_2$ terminated configuration can exist in either a magnetic or multiferroic ground state depending on the chemistry of the nanodot. We find that all the nanodots except $NaNb_8O_{12}$ assume multiferroic order and the isolation mechanism prevalent in other known bulk multiferroics explains the magneto-electric coupling observed in these nanodots. We therefore predict that the suppression of ferroelectricity is not intrinsic in the nanodot geometry.

We also propose a structural factor called the cluster tolerance factor (CTF) via which one can predict the crystallographic nature of the ground state at the smallest of scales. Comparing B-O and A-O bond lengths in both nanodot terminations to the sum of ionic radii revealed that B-O hybridization plays a key role in the stabilization of non-cubic structures in B-O2 terminated nanodots, similar to what is observed in bulk perovskites. Bulk-like polarization is predicted for the $KNb_8O_{12}$, $SrTi_8O_{12}$, $BaZr_8O_{12}$, $BaHf_8O_{12}$, $BaTi_8O_{12}$, and $CaTi_8O_{12}$ nanodots. These results indicate a complex interplay between material chemistry and surface termination in stabilizing both ferroelectricity, magnetism, and multiferroicity at the limit of miniaturization.

Continuing ab initio molecular dynamics simulations will provide insight into whether such non-polar to polar and non-polar to multiferroic transformations can be realized at finite temperatures. Further, computational stability of ferroelectric, magnetic, and/or multiferroic nanodots in solution, on surfaces, or embedded within a solid matrix will provide additional indication of experimental viability. We envisage this work providing plausible pathways towards engineering highly desirable non-volatile memories (FeRAMs) with densities approaching several Tb/in$^2$ and beyond,[45] as well as devices for nanoscale actuating applications and further nanoscale multiferroic memories with electrical write and magnetic read operations.[21]

## Data Availability

The authors declare that all data discussed and analyzed in the current study have been summarized in the manuscript and the supplementary material and will be provided upon reasonable request.

## Acknowledgements

This work was supported by the US Air Force Research Laboratory, Grant No. FA9451-16-1-0040. Part of this work was performed under the auspices of the U.S. Department of Energy by Lawrence Livermore National Laboratory under Contract DE-AC52-07NA27344. Computational resources of nanoHUB and Purdue University are gratefully acknowledged.

## Author Contributions

K.G.V and A.S conceived the research project and designed the necessary simulations. K.G.V conducted all the simulations and performed necessary analysis with help from S.T.R and A.S. All authors contributed equally to manuscript preparation.

## Competing Interests

The authors declare no competing interests.

# Supplementary material for "Ferro-electricity in free standing perovskite-based nanodots: A density functional theory study"

Karthik Guda Vishnu,[1] Samuel Temple Reeve[2], and Alejandro Strachan[1*]

[1]School of Materials Engineering and Birck Nanotechnology Center, Purdue University, West Lafayette, Indiana 47906

[2]Materials Science Division, Lawrence Livermore National Laboratory, Livermore, California, USA 94550

A detailed discussion of results obtained from the SeqQuest code,[i,ii,iii] (PBE-SeqQuest) are discussed in each of the sections below.

## Section S1. Comparison between dipole moments obtained using partial atomic charges (Mullikan) and from integration of electron densities for the ground state structures obtained using PBE-SeqQuest method

Partial atomic charges are obtained on all atoms in the relaxed atomic configurations using Mullikan population analysis[iv] and the total dipole moment ($\vec{p}$) of the nanodots is calculated using the equation below, as the nanodots remain charge neutral:

$$\vec{p} = \sum_{i}^{N} q_i \cdot \vec{r}_i$$

where $\vec{r}_i$ and $q_i$ are the position vector and partial atomic charge on atom $i$, respectively, and N is the total number of atoms.

From Figure S1, we can see that the partial atomic charge method in some cases agree with the electron density integration, but often overestimates the dipole moment, except in one case for $SrTi_8O_{12}$.

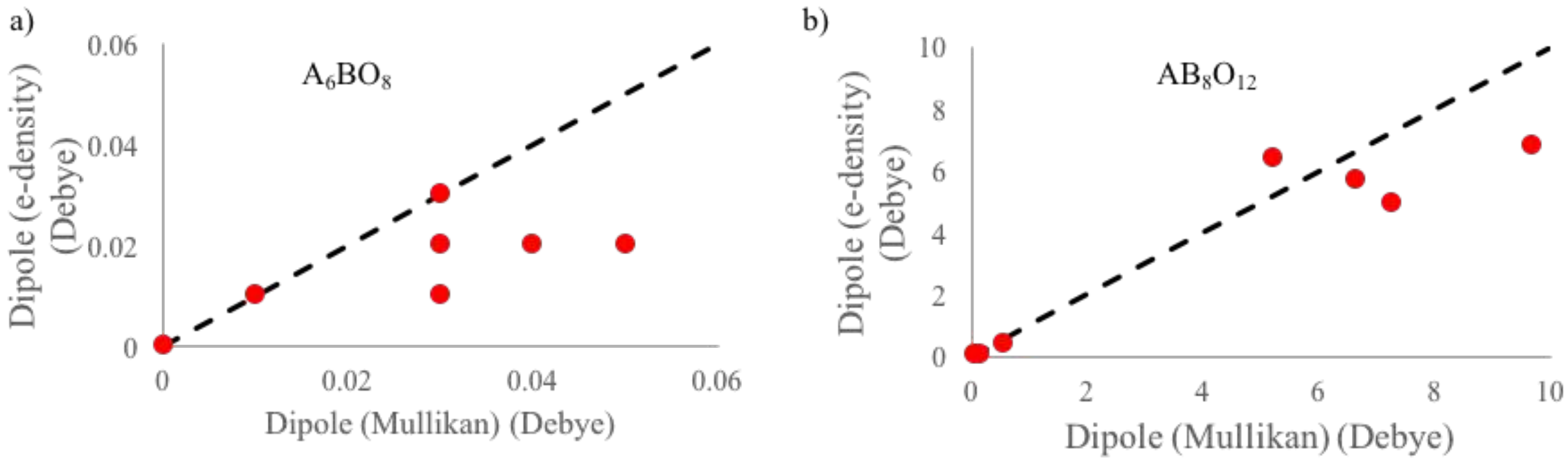

Figure S1: Comparison between dipole moments obtained using partial atomic charges (Mullikan) and integration of electron density for ground state a) A-O terminated nanodots and b) B-$O_2$ terminated nanodots

## Section S2. Symmetry and electrical polarization of relaxed nanodot structures obtained using PBE-SeqQuest method

For all A-O terminated nanodots, the ground state is a cubic structure with $O_h$ point group. Furthermore, they all exhibit large magnetic moments except for A=Na and K and B=Nb. As with the B3LYP-ORCA method, we find that ferroelectricity is completely suppressed at the limit of miniaturization in the A-O terminated nanodots for all chemistries considered.

All B-$O_2$ terminated nanodots, except for $BaTi_8O_{12}$, relax to non-cubic ground state structures and some of these structures exhibit electric dipole moments. These results agree with the B3LYP-ORCA method, except for $BaTi_8O_{12}$, with small differences in space group and exact polarization. Symmetry analysis shows both $NaNb_8O_{12}$ and $CaTi_8O_{12}$ nanodots relax to a non-cubic, centrosymmetric (i.e. non-polar) structure with the point groups $C_i$ and $C_{2h}$ respectively, whose lowest non-vanishing multipole moment is a quadrapole moment. Both $BaHf_8O_{12}$ and $BaZr_8O_{12}$ nanodots relax to $C_{4v}$ (tetragonal) configuration with a preferred axial (z) dipole orientation and a dipole magnitude of 5.69 D and 4.97 D respectively. Ground state structures of $SrTi_8O_{12}$ and $KNb_8O_{12}$ nanodots are polar with point group $C_s$, and they also carry the highest dipole moments (6.38 D and 6.83 D) among all systems investigated. The polarization vector is along the body diagonal of the nanodot ($[111]_{cubic}$) in both structures. Snapshots of all the nanodots that relax to non-cubic ground states are shown in Figure S2.

As with B3LYP-ORCA, it is clear that the existence of a polar ground state in the bulk does not guarantee a polar ground state in the nanodot, as evidenced by $BaTiO_3$ and $NaNbO_3$. Further, the absence a polar bulk phase does not result in suppression of polarization in the nanodots as shown in the case of $SrTiO_3$, $BaZrO_3$ and $BaHfO_3$. Finally, in the case of $KNbO_3$, polarization is stabilized in the both the nanodot geometry and the bulk.

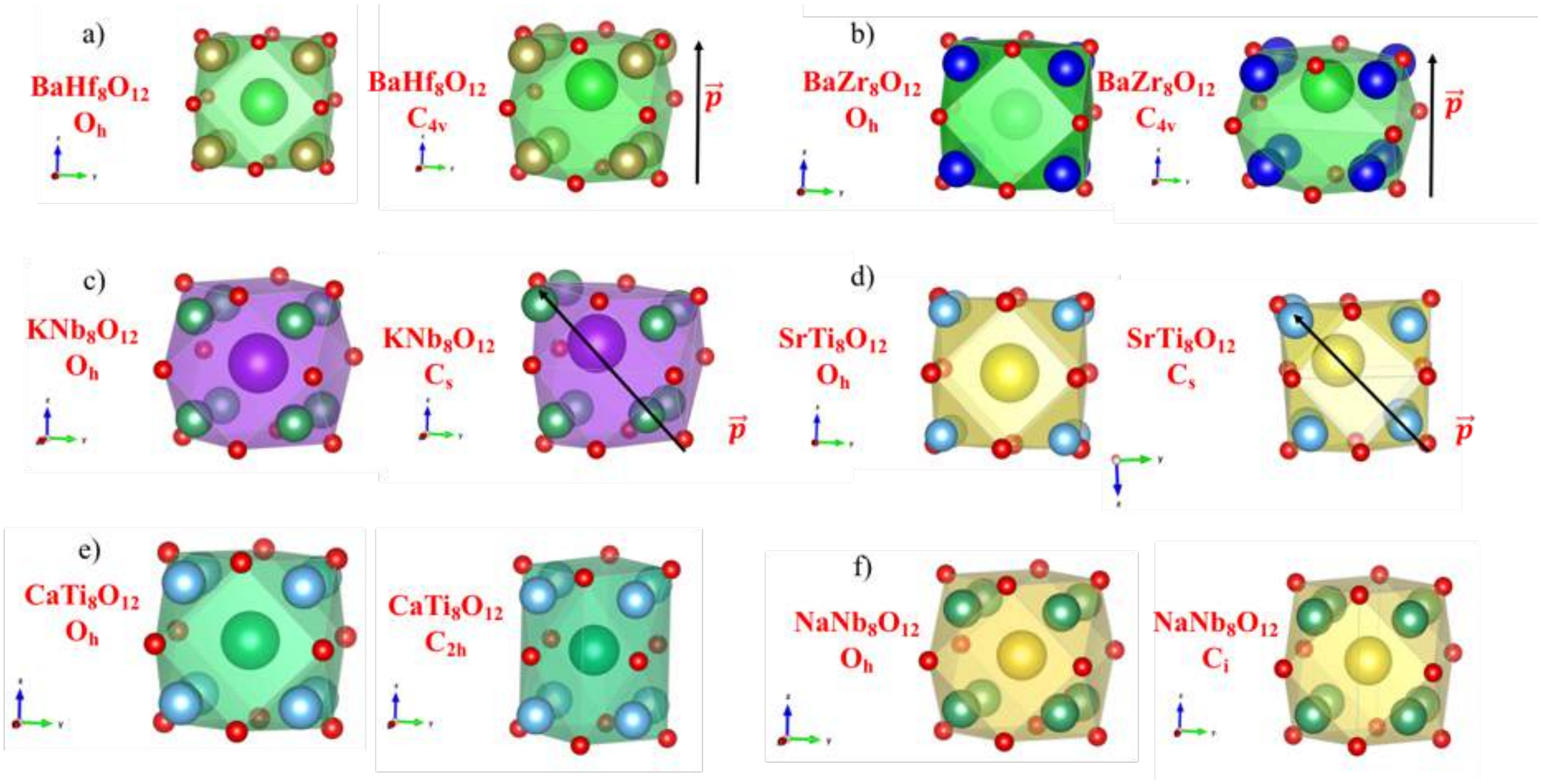

Figure S2: Various nanodots that relax to a non-cubic ground state structure (R) along with their cubic structures (L). A polyhedronaround the central atom is also shown.

## Section S3. Magnetism and presence multiferroic states in relaxed nanodot structures obtained using PBE-SeqQuest method

Figure S3 shows the spin density distribution for the polar (B-$O_2$) nanodots together with those of their corresponding cubic structures. Overall, we find fewer multiferroic B-$O_2$ terminated nanodots since semi-local DFT functionals often underestimate magnetism. In case of $SrTi_8O_{12}$, the polar ground state carries a net spin of 4, thus making it multiferroic. Ferroelectricity is caused by the off-center displacements of the central Sr atom (see Figure S2(d)), while Ti is the maximum atomic spin density carrier, see Figure S3(d). There is, however, a decrease in net magnetic moment from the paraelectric cubic structure (with spin 10) to the polar ground state structure (4). There is also an asymmetry in the net atomic spin density carried by the Ti atoms; those closer to the central Sr atom (bottom left corner of Figure S3(d)) carry the maximum atomic spin density and the atomic spin density on Ti atoms decreases with increasing Sr-Ti bond length. Similarly, in the case of the $KNb_8O_{12}$ multiferroic nanodot (net spin ~5), ferroelectricity is caused by the off-center displacements of the central K atom (see Figure S2(c)), while Nb is the maximum atomic spin density carrier, see Figure S3(c). In both cases, this is similar to the isolation mechanism observed in bulk multiferroics. In the case of $KNb_8O_{12}$, a net increase in magnetic moment from the cubic paraelectric structure (net spin 0.53) to the polar ground state structure (net spin 5.03) which contrasts with $SrTi_8O_{12}$.

In the case of both the $BaHf_8O_{12}$ and $BaZr_8O_{12}$ nanodots, the ground state polar nanodots are non-magnetic. We find that in both these nanodots ferroelectricity is caused by the off-center displacements of the central Ba atom, see Figures S2(a) and S2(b), and that Ba is also the maximum atomic spin density carrier. This shows that the isolation mechanism does not play a role in these nanodots making them only ferroelectric. It should be noted that their non-polar, centrosymmetric cubic counterparts carry a net spin of 2.0.

Spin resolved density of states for all the polar nanodots ($BaHf_8O_{12}$, $BaZr_8O_{12}$, $SrTi_8O_{12}$ and $KNb_8O_{12}$) are also shown in Figure S4. It is clear that in case of both $BaHf_8O_{12}$ and $BaZr_8O_{12}$, spin up (α) and spin down (β) states cancel out each other, resulting in a zero net magnetic moment whereas in case of both $SrTi_8O_{12}$ and $KNb_8O_{12}$, spin up (α) and spin down (β) states do not compensate for each other resulting in a net magnetic moment of 4.0 and 5.03 respectively. Total atomic spin density maps obtained from Mullikan populations of all polar nanodots are also shown Figure S5. The total atomic spin density maps reaffirm that the peripheral B atoms are the majority spin carriers while the off-center displacements of the central A atom lead to polarity.

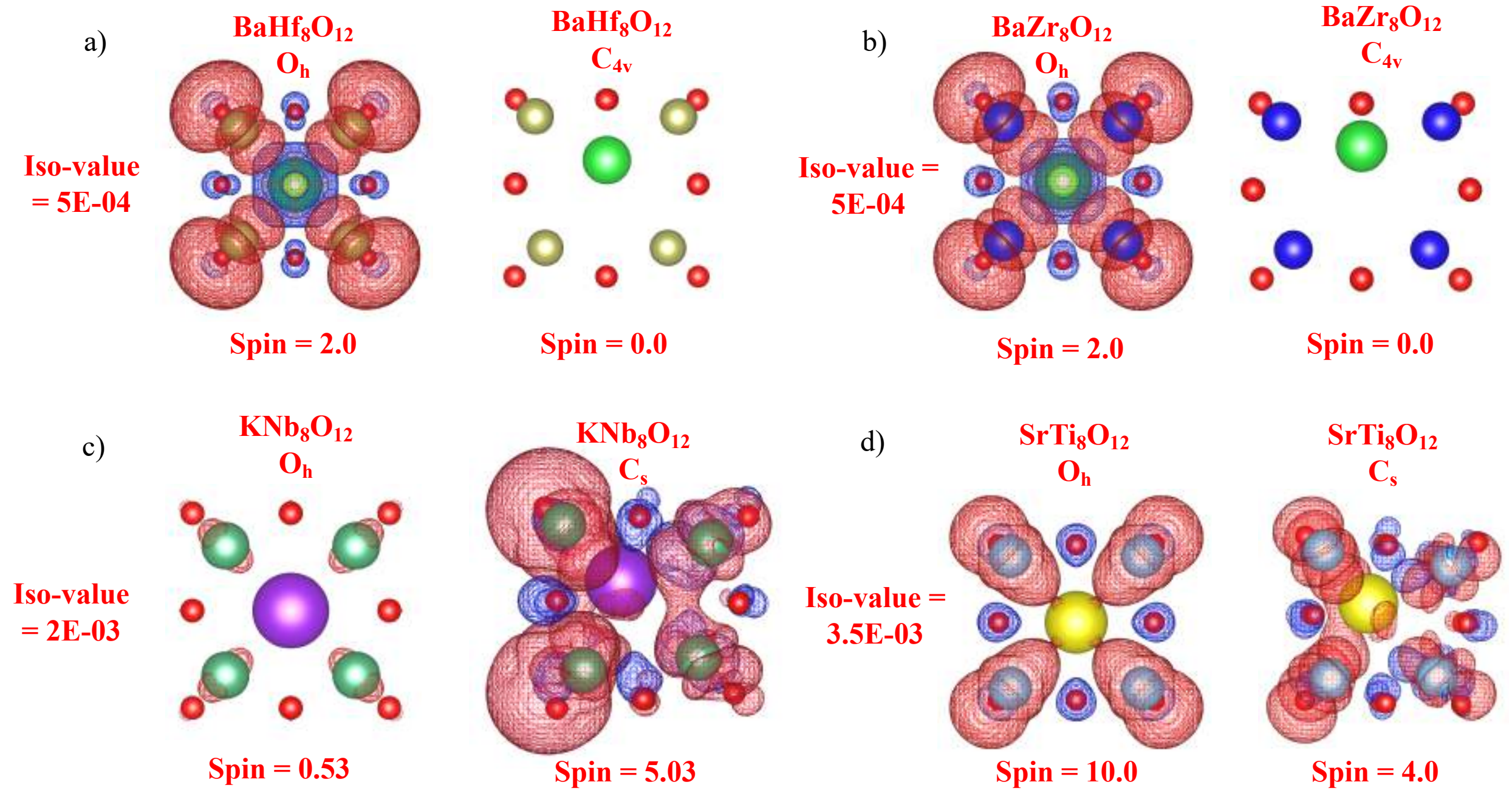

Figure S3: Spin density ($\rho_{up}$ - $\rho_{dn}$) plots for various polar nanodots along with their cubic structures. Iso-surface values corresponding to each plot are also shown. Red and blue colors indicate positive and negative values.

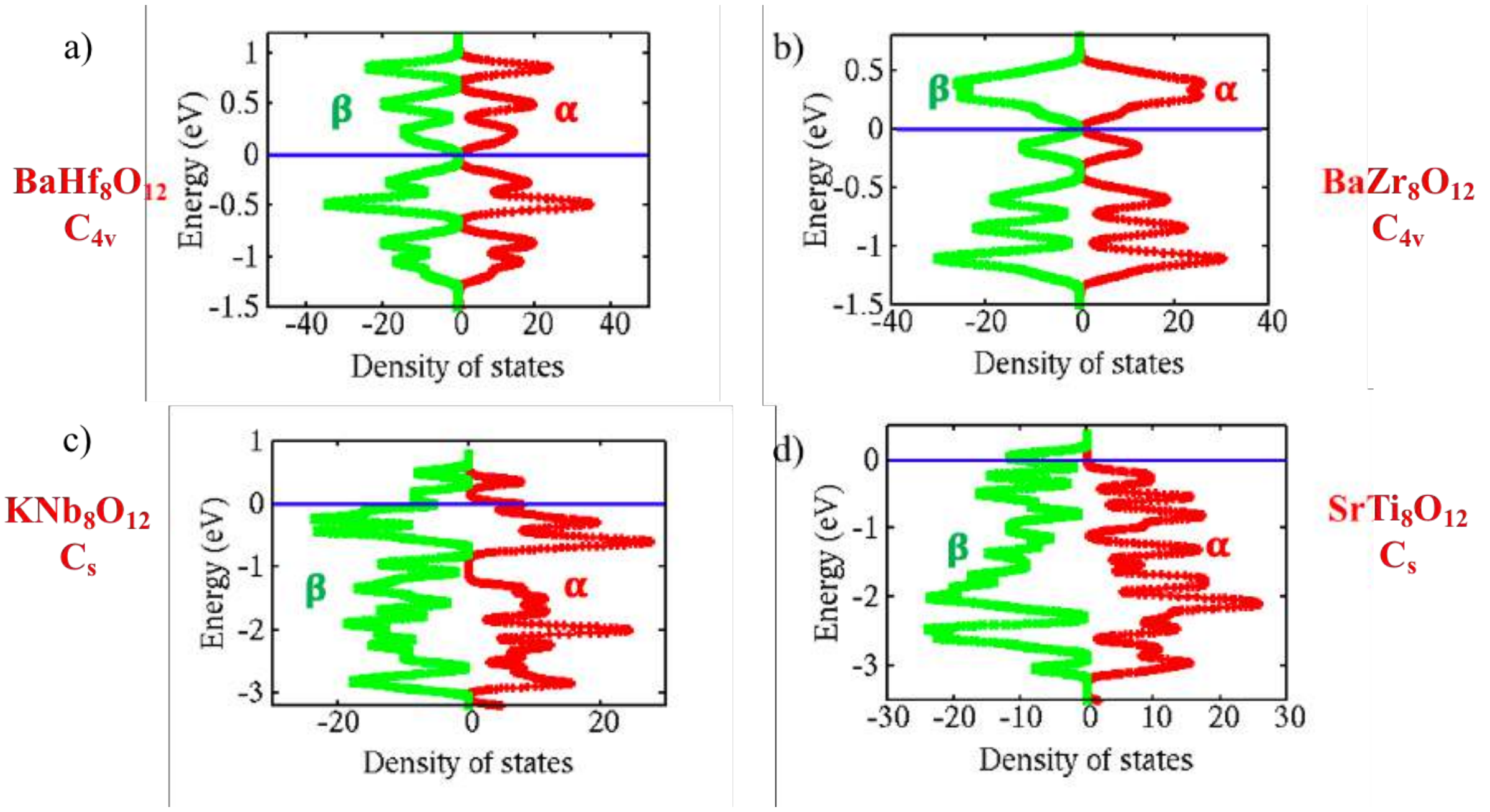

Figure S4: Spin resolved density of states obtained for all the polar nanodots. Energies are scaled with their respective Fermi energy and α denotes "spin up" while β represents "spin down" states. Negative scale on the x-axis is used only to distinguish between the spin up and spin down states.

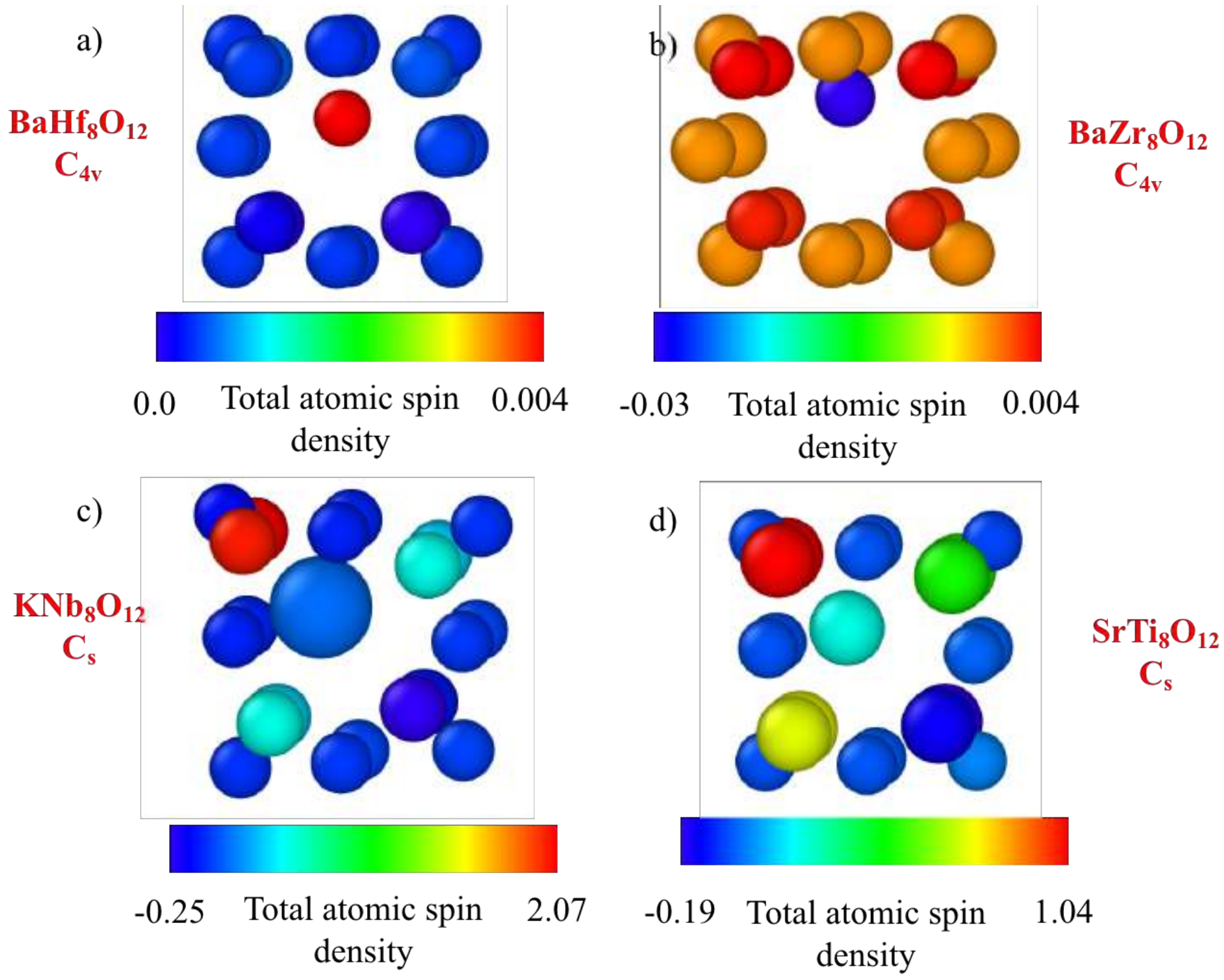

Figure S5: Total atomic spin density maps for the polar nanodots. The color maps are generated using OVITO.[v]

## Section S4. Bond length analysis of cubic centrosymmetric non-polar nanodots obtained using PBE-SeqQuest method

For comparison with the B3LYP-ORCA results, the same bond length analysis was performed for PBE-SeqQuest, as shown in Figure S6. All the comparisons and conclusions from B3LYP-ORCA remain, with minor quantitative changes with PBE-SeqQuest.

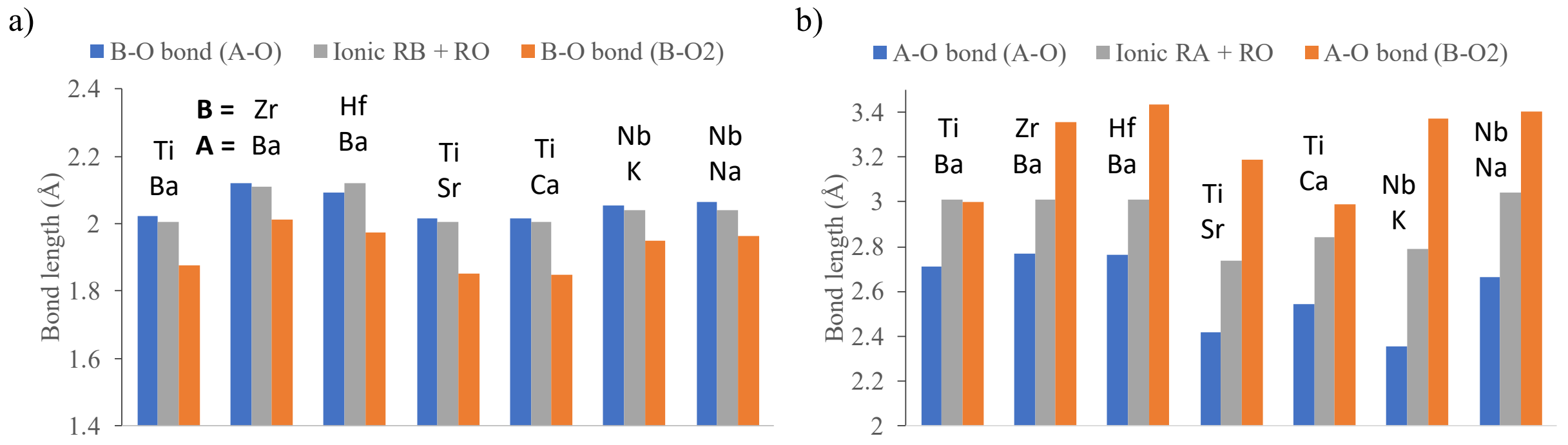

Figure S6: a) Comparison of B-O bond lengths from both cubic centrosymmetric A-O and $BO_2$ terminated nanodots against their sum of ionic radii b) Comparison of A-O bond lengths from both cubic centrosymmetric A-O and $BO_2$ terminated nanodots against their sum of ionic radii